\documentclass[12pt]{iopart} 

\usepackage{graphicx}%
\usepackage{multirow}%
\usepackage{amsthm}%
\usepackage{mathrsfs}%
\usepackage[title]{appendix}%
\usepackage{xcolor}%
\usepackage{textcomp}%
\usepackage{manyfoot}%
\usepackage{booktabs}%
\usepackage{algorithm}%
\usepackage{algorithmicx}%
\usepackage{algpseudocode}%
\usepackage{listings}%
\usepackage{siunitx}

\usepackage{subcaption}
\usepackage{array}

\usepackage{comment}

\begin{document}

\title[]{A prototype neutron-detector array for future deep-underground $s$-process studies}

\author{Thomas Chillery$^{1}$, David Rapagnani$^{2,3}$, Chemseddine Ananna$^{4,5}$, Edoardo D'Amore$^{6,7}$,  Gianluca Imbriani$^{2,3}$, Antonino di Leva$^{2,3}$, Daniela Mercogliano$^{2,3}$, Jakub Skowronski$^{6,7}$, Benjamin Br\"uckner$^{8}$, Sophia Dellmann$^{9}$, Philipp Erbacher$^{8}$, Tanja Heftrich$^{8}$, Ren\'e Reifarth$^{8,9}$, Mario Weigand$^{8}$, Andreas Best$^{2,3}$}

\address{$^{1}$Laboratori Nazionali del Gran Sasso, Istituto Nazionale di Fisica Nucleare, Via G. Acitelli 22, L'Aquila - Localita Assergi, 67100, AQ, Italy}

\address{$^{2}$Dipartimento di Fisica, Università degli Studi di Napoli ``Federico II”, Via Cintia 21, Napoli, 80126, NA, Italy}

\address{$^{3}$INFN, Sezione di Napoli, Napoli, Italy}

\address{$^{4}$Dipartimento di Matematica e Fisica, Università del Salento, Via per Arnesano snc, Lecce, 73100, LE, Italy}

\address{$^{5}$INFN, Sezione di Lecce, Lecce, Italy}

\address{$^{6}$Dipartimento di Fisica, Università degli Studi di Padova, Via Marzolo 8, 35131 Padova, Italy}

\address{$^{7}$INFN, Sezione di Padova, 35131 Padova, Italy}

\address{$^{8}$Goethe University Frankfurt, Max-von-Laue Straße 1, 60438 Frankfurt, Germany}

\address{$^{9}$Los Alamos National Laboratory, 87545 Los Alamos, NM, USA}

\ead{thomas.chillery@lngs.infn.it}

\begin{abstract}
We report a novel neutron-detection approach employing an EJ-309 liquid scintillator surrounded by six $^{3}$He proportional counters. Tests were performed at the FRANZ facility of the Goethe-University Frankfurt using the $^{7}$Li(p,n$_{0}$)$^{7}$Be reaction, producing neutrons across energies 50 -- 720~keV. The scintillator's neutron energy quenching is determined, and its neutron/$\gamma$-ray discrimination performance is evaluated. The lowest detectable neutron energy is 163~keV. EJ-309 - $^{3}$He counter neutron coincidences are compared with those from simulations. This array forms the prototype of a larger design, called SHADES, currently undergoing construction and testing for an upcoming deep-underground study of the $^{22}$Ne($\alpha$,n)$^{25}$Mg reaction cross-section at the freshly-commissioned Bellotti Ion Beam facility of the INFN Laboratori Nazionali del Gran Sasso. This upcoming project is expected to achieve exceptionally low sensitivity for measuring the cross section at energies of interest for the astrophysical ``weak" and ``main" slow neutron-capture processes.
\end{abstract}

\section{Introduction} \label{sec:introduction}

Nuclear reactions involving the capture or emission of neutrons play a leading role in several astrophysical scenarios. In particular, the slow neutron-capture ($s$-) process is widely regarded to occur in massive stars (8M$_{\odot}$) and asymptotic giant branch (AGB) stars, synthesizing neutron-rich isotopes above A $\sim$ 60 -- 90 and 90 -- 209, respectively~\cite{Pignatari_2010, Bisterzo_2015}. The nucleosynthesis flow is predominantly fueled by the neutron sources $^{13}$C($\alpha$,n)$^{16}$O and $^{22}$Ne($\alpha$,n)$^{25}$Mg. The $^{13}$C($\alpha$,n)$^{16}$O reaction has seen intense interest~\cite{Cristallo_2018, Czedreki_2024}, particularly in recent years with sensitive underground experiments~\cite{Ciani_2021, Gao_2022} pushing cross-section measurements into the astrophysical energy regime. However, the $^{22}$Ne($\alpha$,n)$^{25}$Mg reaction cross-section remains weakly constrained at astrophysical temperatures of interest~\cite{Adsley_2021, Wiescher_2023, Best_2025}; 100 -- 300\,MK. To address this, an experimental campaign is underway at the new Bellotti Ion Beam facility (IBF)~\cite{Sen_2019, Junker_2023} of the INFN Laboratori Nazionali del Gran Sasso (LNGS) to directly measure the cross section at energies of interest, $E_{\mathrm{lab}}^{\alpha}$ = 600~--~886\,keV. A new neutron detector array, titled SHADES (Scintillator-$^{3}$He Array for Deep-underground Experiments on the $S$-process), is planned to measure the neutrons of interest directly.

Since this measurement can be severely impacted by beam-induced background neutrons, some energy sensitivity for the detected neutrons is desirable, in addition to a reasonably high neutron-detection efficiency. For SHADES, a combination of twelve EJ-309 liquid scintillators~\cite{Annana_2024} and eighteen $^{3}$He proportional counters has been selected\footnote{EJ-309 possesses good scintillation and pulse shape discrimination properties while at the same time exhibiting a high flash point, low vapor pressure, and low chemical toxicity, allowing its use in environmentally sensitive locations like the protected area of the LNGS.}. Before constructing the full array, a scaled-down prototype consisting of a single scintillator and six $^{3}$He counters was tested using a tunable-energy neutron beam. The design principle is as follows: neutrons first enter the EJ-309 and deposit the majority of their energy through energy loss events with the scintillator's hydrogen atoms. The neutrons are thereby thermalised (E $<$ 0.1\,eV) before escaping the crystal and entering one of the surrounding $^{3}$He counters, where the thermal neutron is absorbed by the $^{3}$He(n,p)t reaction (Q = 763\,keV). The EJ-309 provides energy, timing, and tagging information for the neutron, and the counters provide a measure of the number of neutrons thermalised in the system. The motivation behind constructing the prototype array is to test this detection principle, which is then expected to be applied to the full SHADES array. Specifically, using the EJ-309 scintillators as active neutron thermalisers and energy detectors, and the $^{3}$He counters to extract measurement yields of the $^{22}$Ne($\alpha$,n)$^{25}$Mg reaction. Through coincidences, the EJ-309 is also planned to act as a veto for beam-induced background events impacting the counters, thereby improving the sensitivity of SHADES. 

This paper reports the performance of this prototype array, with comparisons to simulations to check timing coincidences. Particular emphasis is placed on the neutron/$\gamma$-ray discrimination as a function of neutron energy, the low-energy threshold, and the coincidence characteristics of the combined setup. The structure is as follows. Section \ref{sec:method} summarizes the experimental setup, section \ref{sec:results} highlights the EJ-309 characterization, including a comparison between traditional and machine-learning pulse-shape discrimination (PSD) approaches. Section \ref{sec:coincidences} compares the EJ-309 - counter coincidence time between measurement and a \textsc{Geant4}~\cite{Geant4} simulation. Conclusions and future perspectives for the full SHADES array are provided in section \ref{sec:conclusions}. 

\section{Experimental Method}\label{sec:method}

The experiment was performed at the Goethe University Frankfurt's Van de Graaff accelerator (VdG) facility ``FRANZ"~\cite{Alzubaidi_2016}. The VdG accelerated a proton beam with nominal\footnote{The target chamber current measurement was not electron suppressed.} currents 35 -- 250\,nA at select energies between 1900~--~2450\,keV in approximately 50\,keV steps. The beam bombarded a \SI{3.1}{\micro\meter} thick $^{7}$Li target deposited on copper, producing neutrons between 50~--~720\,keV via the $^{7}$Li(p,n$_{0}$)$^{7}$Be reaction. A summary of the beam runs is provided in Table \ref{tab:runInfo}. 

\begin{figure}[ht]
    \centering
    \begin{subfigure}[ht]{\textwidth}
        \includegraphics[width=\textwidth]{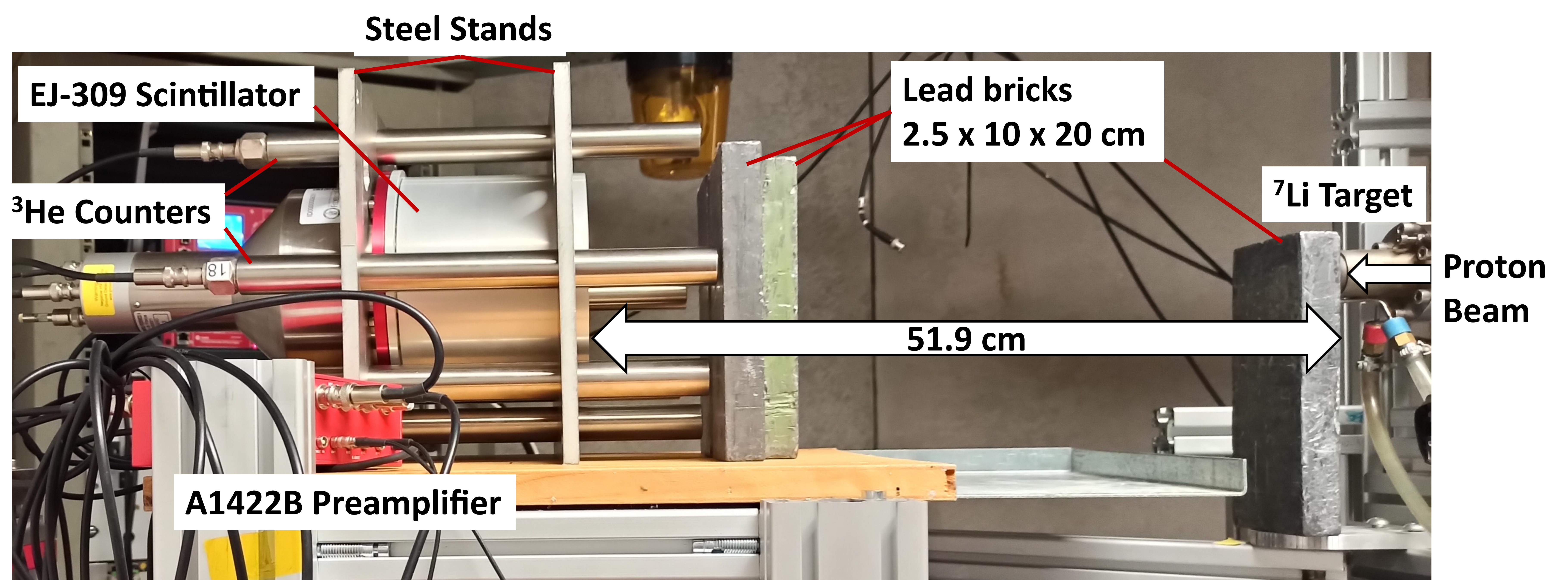}
        \caption{Photograph of the setup installed at ``FRANZ".}\label{fig:setup_a}
    \end{subfigure}
    \begin{subfigure}[hb]{\textwidth}
        \includegraphics[width=0.9\textwidth]{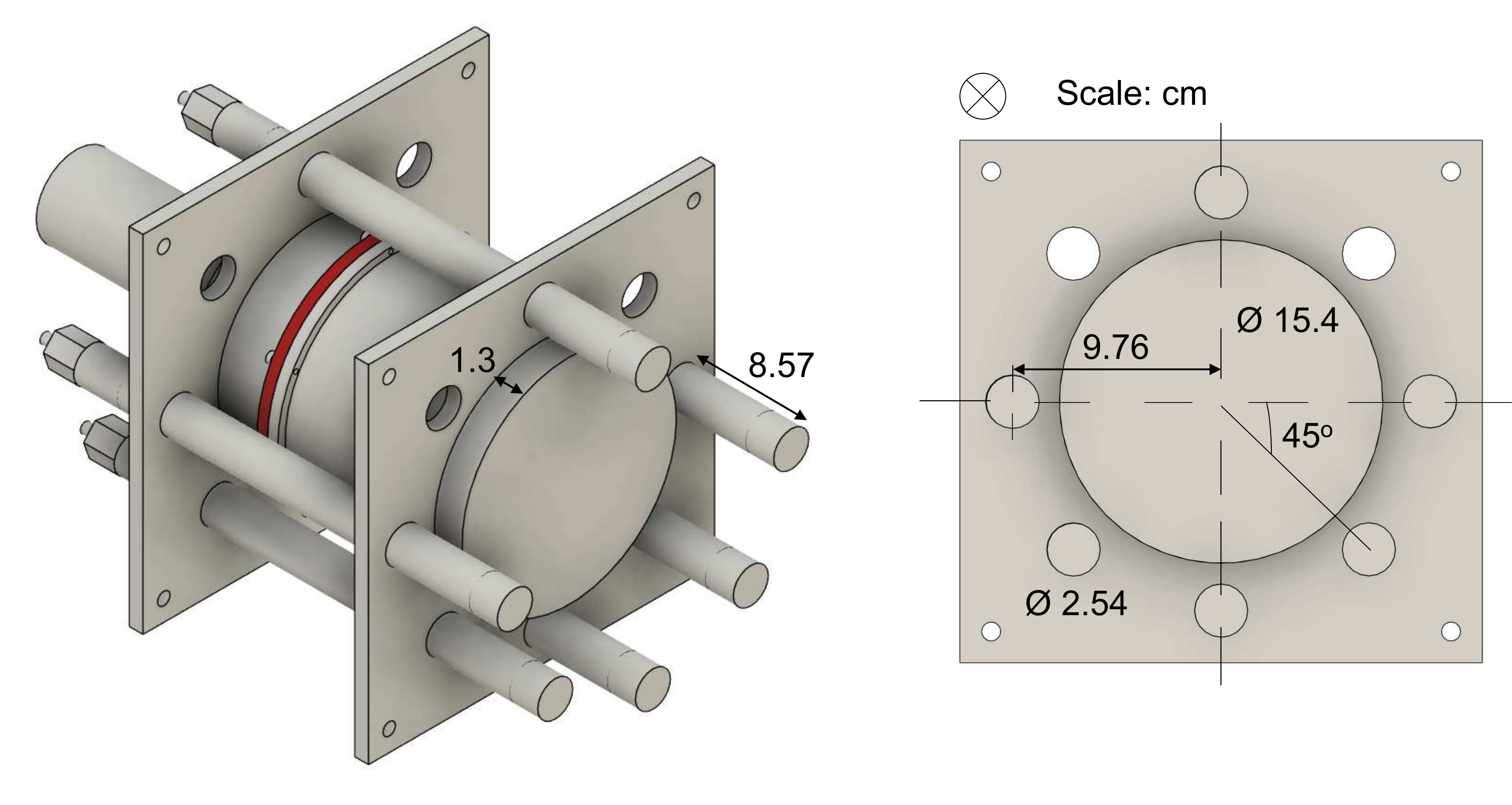}
        \caption{Computer-assisted drawing of the prototype array.}\label{fig:setup_b}
    \end{subfigure}
    \caption{Experimental setup}
    \label{fig:setup}
\end{figure}

\begin{figure}[ht]
\centering
\includegraphics[width=0.8\textwidth]{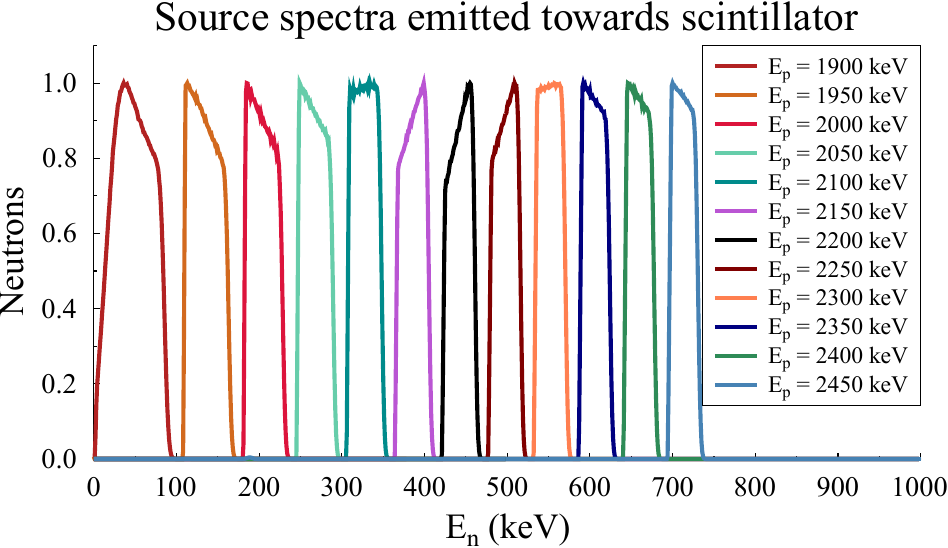}
\caption{Neutron spectra calculated from PINO for the proton energies used in this measurement. (Refer to online plots for color).}\label{fig:pino}
\end{figure}

The prototype setup consisted of a single 12.7 x 12.7\,cm EJ-309 liquid scintillator cylindrically surrounded by six $^{3}$He proportional counters\footnote{GE Reuter Stokes model RS-P4-0810-250, 10\,bar filling pressure}, all supported by two parallel steel plates. The EJ-309 covered the lithium target with a solid angle of ca. 0.07\,sr (8.4$^{\circ}$ cone aperture). A photograph of the setup is provided in Figure~\ref{fig:setup}. The center-to-center distance between the scintillator and counters was 9.76\,cm. The target -- scintillator distance was set to the furthest possible, 51.9\,cm, to minimize the angular exposure and energy dispersion of neutrons reaching the detector. The scintillators front face covered a solid angle 8.4$^{\circ}$ in the laboratory frame. To reduce background from $\gamma$-rays emitted from the target, three lead bricks of 2.5 x 10 x 20\,cm were placed between the target and array. The setup remained otherwise unshielded. The source-like neutron spectra simulated from Monte-Carlo tool ``PINO"~\cite{Reifarth_2009} are provided in Figure \ref{fig:pino}.

\begin{table}[ht]
\centering
\caption{Run information. $E_{\rm{p}}$ is the proton beam energy. The average neutron energy $E_{\mathrm{n}}$ is provided from PINO simulations~\cite{Reifarth_2009} with and without the 7.5\,cm lead bricks used in this experiment. The nominal proton beam current and deposited charge on the $^{7}$Li target are also provided.}\label{tab:runInfo}%
    \begin{tabular}{ 
        | c  
        | c  
        | c 
        | c  
        | c 
        |
        }
    \hline
    \multirow{2}{*}{\centering $E_{\mathrm{p}}$ [keV]} & \multicolumn{2}{c|}{Average $E_{\mathrm{n}}$ [keV]} & \multirow{2}{0.24\textwidth}{\centering Nominal beam current$^{b}$ [nA]} & \multirow{2}{0.26\textwidth}{\centering Deposited charge [\SI{}{\micro\coulomb}]} \\
    \cline{2-3}
    & no lead & 7.5\,cm lead & & \\
    \hline
    2450(2)$^{a}$ & 713 & 690 & 35 & 124 \\
    \hline
    2449(2) & 713 & 690 & 200 & 251 \\
    \hline
    2400(2) & 659 & 642 & 200 & 1121 \\
    \hline
    2349(2) & 605 & 589 & 200 & 901 \\
    \hline
    2299(2) & 552 & 535 & 200 & 892 \\
    \hline
    2249(2) & 497 & 485 & 200 & 400 \\
    \hline
    2200(2)$^{a}$ & 440 & 423 & 35 & 77 \\
    \hline
    2198(2) & 440 & 423 & 150 & 375 \\
    \hline
    2150(2) & 380 & 356 & 150 & 564 \\
    \hline
    2099(2) & 321 & 295 & 250 & 825 \\
    \hline
    2050(2) & 262 & 235 & 150 & 504 \\
    \hline
    2000(2) & 202 & 180 & 150 & 495 \\
    \hline
    1950(2) & 134 & 117 & 160 & 290 \\
    \hline
    1900(2)$^{a}$ & 51 & 47 & 30 & 105 \\
    \hline
    \end{tabular} \\
\vspace{0.5em}
$^{a}$Low beam current runs with reduced EJ-309 energy threshold.
$^{b}$Unsuppressed.
\end{table}

The scintillator photo-multiplier tube\footnote{The EJ-309 was equipped with a 10-stage ETL9390 photo-multiplier tube} was biased to -1025\,V. The counters were connected to two 4-channel CAEN A1422B preamplifiers and biased to +750\,V each. An eight-channel, 14-bit DT5725B CAEN module running the DPP-PSD firmware digitized the waveforms from the detectors at 250\,MSamples/s. A PC running the CAEN CoMPASS DAQ software was used to write waveforms, timestamps, board number, channel, and event flags to disk. Aside from beam measurements, data were collected using $\gamma$-ray calibration sources $^{137}$Cs (Activity: 2.43\,kBq) and $^{60}$Co (Activity: $<$1\,kBq) for calibration of the scintillator energy signals.

\section{Scintillator Performance: Calibration and Pulse Shape Analysis}\label{sec:results}

\subsection{EJ-309 Energy Calibration}

\begin{figure}[ht]
\centering
\includegraphics[width=\textwidth]{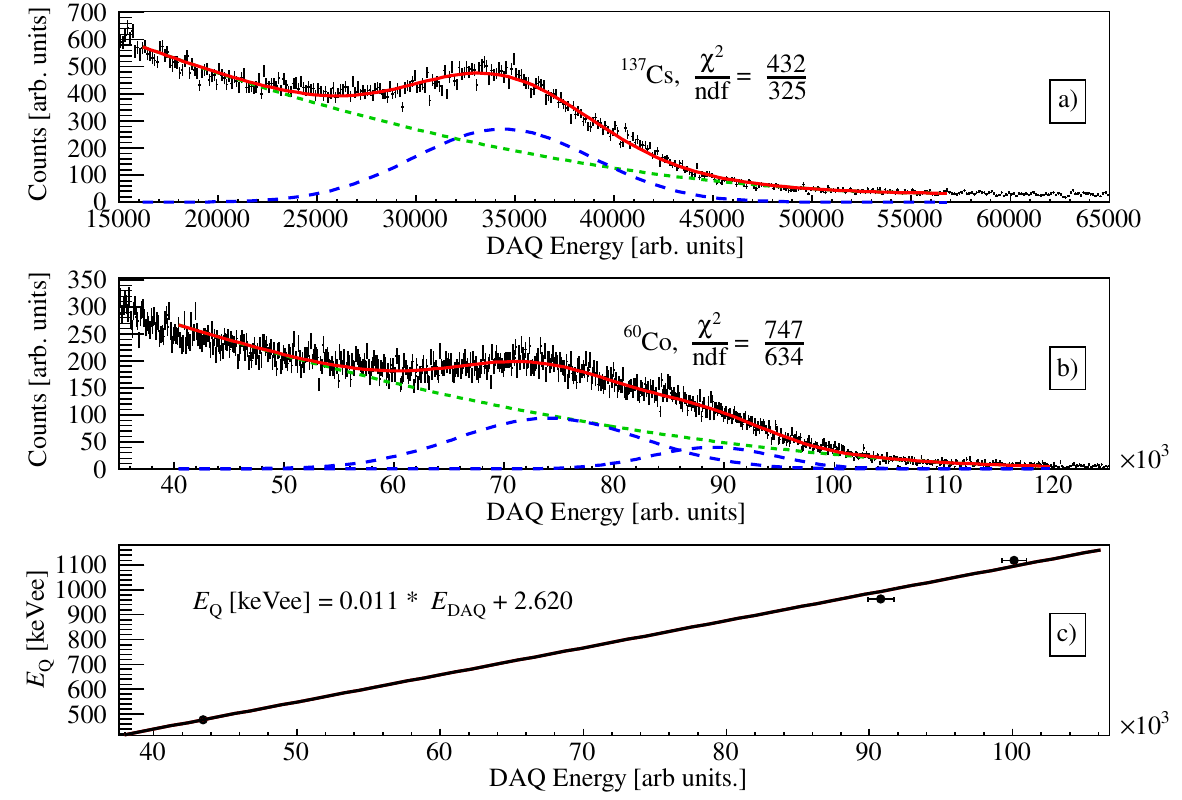}
\caption{Energy spectra measured by the EJ-309 for fixed $\gamma$-ray sources a) $^{137}$Cs and b) $^{60}$Co, zoomed into the Compton edges. Solid red lines represent the total fit, with quoted $\chi^{2}$/ndf. Dotted green lines represent the background modeled by a 2$^{\mathrm{nd}}$ order polynomial (omitting the peak region of interest). Dashed blue lines represent only the Gaussian component of the total fit. c) DAQ energy to quenched energy ($E_{\mathrm{Q}}$) calibration. (Refer to online plots for color).}\label{fig:scintSourceFits}
\end{figure}

The EJ-309 scintillator was energy calibrated using Compton edges of the $\gamma$-rays emitted by the sources $^{137}$Cs and $^{60}$Co. Fits and corresponding quenched-energy calibration, in units of electron-equivalent energy (MeVee)~\cite{Knoll_2000}, are shown in Figure~\ref{fig:scintSourceFits}. 

\begin{figure}[ht]
\centering
\includegraphics[width=\textwidth]{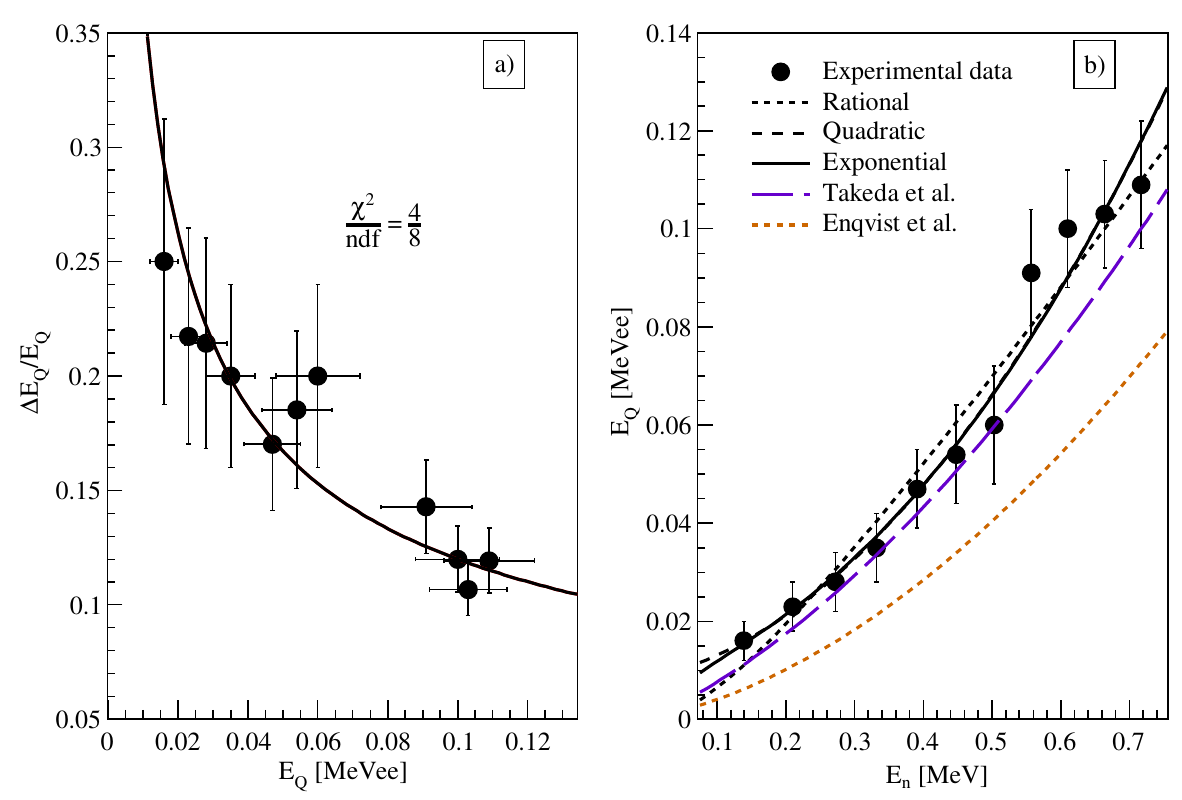}
\caption{a) Energy resolution of the EJ-309 scintillator as a function of quenched energy. b) Quenched energy to neutron energy calibration (points). Three different fits are shown: a rational function (short-dashed), a quadratic (long-dashed) and an exponential function (solid). Literature examples are also plotted from Takeda~\cite{Takeda_2011} (purple dashed) and Enqvist~\cite{Enqvist_2013} (orange short-dashed), for the same crystal size as this study. (Refer to online plots for color).}\label{fig:scintEqEn}
\end{figure}

The quenched energy resolution, $\Delta E_{\mathrm{Q}}/E_{\mathrm{Q}}$, is plotted as a function of $E_{\mathrm{Q}}$ in Figure~\ref{fig:scintEqEn}a and the resolution was fitted with the equation~\cite{Kornilov_2009, Enqvist_2013}:
\begin{equation}
    \frac{\Delta E_{\mathrm{Q}}}{E_{\mathrm{Q}}} = \sqrt{ \alpha^{2} + \frac{\beta^{2}}{E_{\mathrm{Q}}} + \frac{\gamma^{2}}{E_{\mathrm{Q}}^{2}} },
\end{equation}
where the present fit has parameters $\alpha$~=~0.03(8), $\beta$~=~0.037(5)\,$\sqrt{\rm{MeVee}}$, and $\gamma$~=~0.000(3)\,MeVee. 
The quenched energy to neutron energy calibration is plotted in Figure~\ref{fig:scintEqEn}b, where the neutron energy was calculated kinematically~\cite{maryon_young} considering the incident proton energy and the emitted neutron angle $\theta_{\mathrm{lab}}$ = 4.3$^{\circ}$. The data are fitted with three different models: a rational function (short-dashed), a quadratic function (long-dashed), and an exponential (solid). Additionally, the present data is compared with exponential fits from previous EJ-309 quenching studies~\cite{Enqvist_2013,Takeda_2011}. The present data is in good agreement with the exponential function from~\cite{Takeda_2011}, yet shows disagreements of 2 -- 3 sigma to the exponential function from~\cite{Enqvist_2013}. Given the literature data were collected at higher energies to those of this study, the difference may be due to the extrapolation below 0.7~MeV. Also, the quenching effect is dependent upon the light collection properties of unique crystals. It's likely the growing / construction techniques of the EJ-309 crystals combined with different PMTs used across these studies cause differences in the quenching factor of neutrons. 

Present fits of the data in Figure~\ref{fig:scintEqEn} are defined as follows: \\
Rational:
\begin{equation}
    E_{\mathrm{Q}}(E_{\mathrm{n}}) = \frac{P_{0} \cdot E_{\mathrm{n}}^{2}}{E_{\mathrm{n}} + P_{1}}
\end{equation}
Quadratic:
\begin{equation}
    E_{\mathrm{Q}}(E_{\mathrm{n}}) = P_{0} \cdot E_{\mathrm{n}}^{2} + P_{1} \cdot E_{\mathrm{n}} + P_{2}
\end{equation}
Exponential:
\begin{equation}
    E_{\mathrm{Q}}(E_{\mathrm{n}}) = P_{0} \cdot E_{\mathrm{n}} - P_{1} \cdot [1 - \exp(-P_{2} \cdot E_{\mathrm{n}}^{P_{3}} )]
\end{equation}
The fit parameters are provided in table \ref{tab:EqEnPars} with one sigma uncertainties. The quadratic and exponential functions best describe the data across this low neutron energy range. The quadratic formula, which has a mildy better chi$^2$/ndf (requires one fewer parameter) to the exponential fit, was thus applied for subsequent EJ-309 energy calibration of the runs collected with beam. Following energy calibration, the EJ-309 PSD capabilities were evaluated using both traditional and machine-learning methods. 

\begin{table}[ht]
\centering
\caption{$E_{\mathrm{Q}}$ vs $E_{\mathrm{n}}$ fit parameters for the black curves in Figure~\ref{fig:scintEqEn}b.}\label{tab:EqEnPars}%
\begin{tabular}{| c | c | c | c | c | c | }
\hline
Fit & $\chi^{2}/\mathrm{ndf}$ & $P_{0}$ & $P_{1}$ & $P_{2}$ & $P_{3}$ \\
\hline
Rational    & 5.31/9 & 0.20(4)  & 0.20(13) & -         & - \\
\hline
Quadratic   & 2.62/8 & 0.17(8)  & 0.03(6)  & 0.008(10) & -  \\
\hline
Exponential & 2.72/7 & 1.15(8)  & 2.5(2)   & 0.47(4)   & 1.048(17) \\
\hline
\end{tabular}
\end{table}

\subsection{Traditional Pulse Shape Discrimination}

\begin{figure}[ht]
\centering
\includegraphics[width=0.8\textwidth]{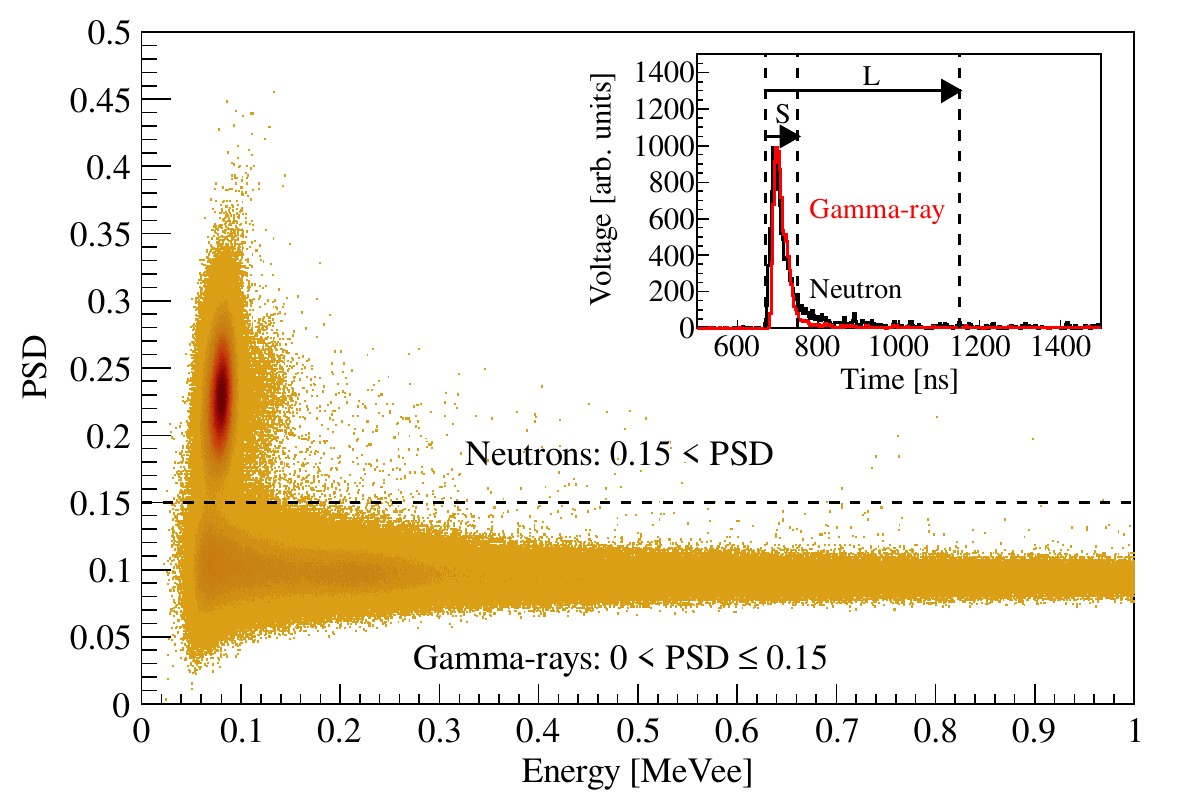}
\caption{Traditional PSD vs quenched energy. Proton beam energy = 2449\,keV. Inset: Sample waveforms for neutron (PSD = 0.21) and $\gamma$-ray (PSD = 0.09). The long and short integral ranges are also shown. (Refer to online plots for color).}\label{fig:psdVsEnergy}
\end{figure}

\begin{figure}[ht]
\centering
\includegraphics[width=\textwidth]{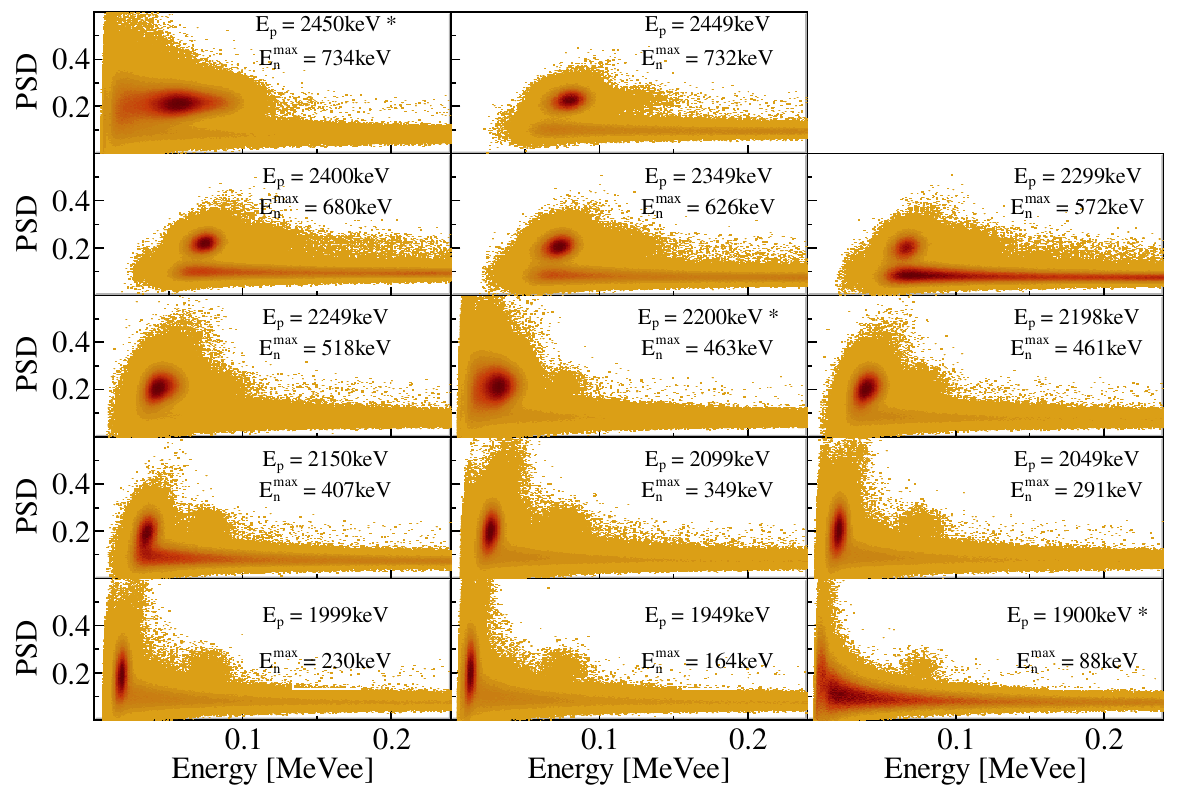}
\caption{Traditional PSD vs quenched energy. Labels indicate the incident proton energy and maximum neutron energy, both in keV. The three runs with labels marked by * (2450\,keV, 2200\,keV, and 1900\,keV) were collected with lower beam intensity and energy threshold settings. (Refer to online plots for color).}\label{fig:psdVsEnergy_All}
\end{figure}

The scintillator waveforms were first pre-processed by evaluating and subtracting their baseline, disregarding any waveforms affected by pileup or flagged by the DAQ as saturated. It was found the CoMPASS DAQ flags were not comprehensive in removing all pileup events, and thus a manual peak-finding approach similar to that of~\cite{Luo_2018} was performed. Peak positions were found by taking the derivative of the waveform and selecting those above both a certain derivative threshold and amplitude threshold of the original signal. Only scintillator events with single peaks were used in the subsequent PSD analysis. To obtain the PSD values the signals were integrated between two gates, a short (S) and long (L), and the PSD parameter was calculated as PSD = 1 - S/L. The gate ranges were optimized~\cite{Chemseddine_thesis} to 80\,ns for the short and 480\,ns for the long one. Given the same amplitude, neutron signals have a longer tail than $\gamma$-ray signals and therefore have a larger PSD value. Figure~\ref{fig:psdVsEnergy} shows the PSD vs quenched energy for a proton beam energy of 2449\,keV (max neutron energy 732\,keV). Sample waveforms for a neutron and $\gamma$-ray event are shown in the figures inset. Neutrons have a PSD parameter above 0.15, whereas $\gamma$-rays fall in the range of 0.01 -- 0.15. The neutrons appear as a locus because their low energy is almost fully deposited in the scintillator, whereas the $\gamma$-rays cover a broader energy range since they arise from natural background sources and beam interactions. PSD vs quenched energy plots for all the beam energies used in this study are plotted in Figure~\ref{fig:psdVsEnergy_All}.

From Figure~\ref{fig:psdVsEnergy_All}, there is a clear locus emerging around 0.07 -- 0.1\,MeVee, particularly at beam energies below 2249\,keV. This arises from boron contamination in the scintillator crystal via alphas emitted through the $^{10}$B(n,$\alpha$)$^{7}$Li reaction. It is also apparent the traditional PSD method faces a low energy limit at approximate proton energy 2200\,keV (maximum neutron energy 461\,keV), where the neutron and $\gamma$-ray loci start overlapping below energies of 60\,keVee. To push the discrimination limit as low as possible in terms of neutron energies, a novel approach using artificial neural networks was developed and is described in the following section. 

\subsection{PSD - Neural Network}

In recent years, neural networks have emerged as powerful tools in nuclear physics, offering enhanced experimental sensitivity for tasks such as particle identification, signal processing, and waveform discrimination in liquid scintillators~\cite{Jinia_2021, Boehnlein_2022, Wang_2022, Abdulaziz_2024}. These networks operate through interconnected layers of artificial neurons, each layer designed to extract hierarchical features from input data. Once the architecture, defined by the number of layers, neurons, and their connectivity, is optimized, the network undergoes training using observed or simulated datasets. During this process, the network iteratively adjusts weights to minimize a predefined loss function, analogous to the classical $\chi ^{2}$ minimization but executed through gradient-based optimization. Critical to this optimization are three hyperparameters: the learning rate, which governs the step size of weight adjustments; the batch size, defined by the subset of data samples used per optimization step; and the number of epochs, corresponding to full passes through the training dataset. A carefully tuned learning rate ensures stable convergence, while the batch size balances computational efficiency and statistical robustness. The epoch count determines the duration of training, preventing underfitting or overfitting. 

\begin{figure}[ht]
    \centering
    \begin{minipage}{0.49\textwidth}
        \centering
        \includegraphics[width=\linewidth]{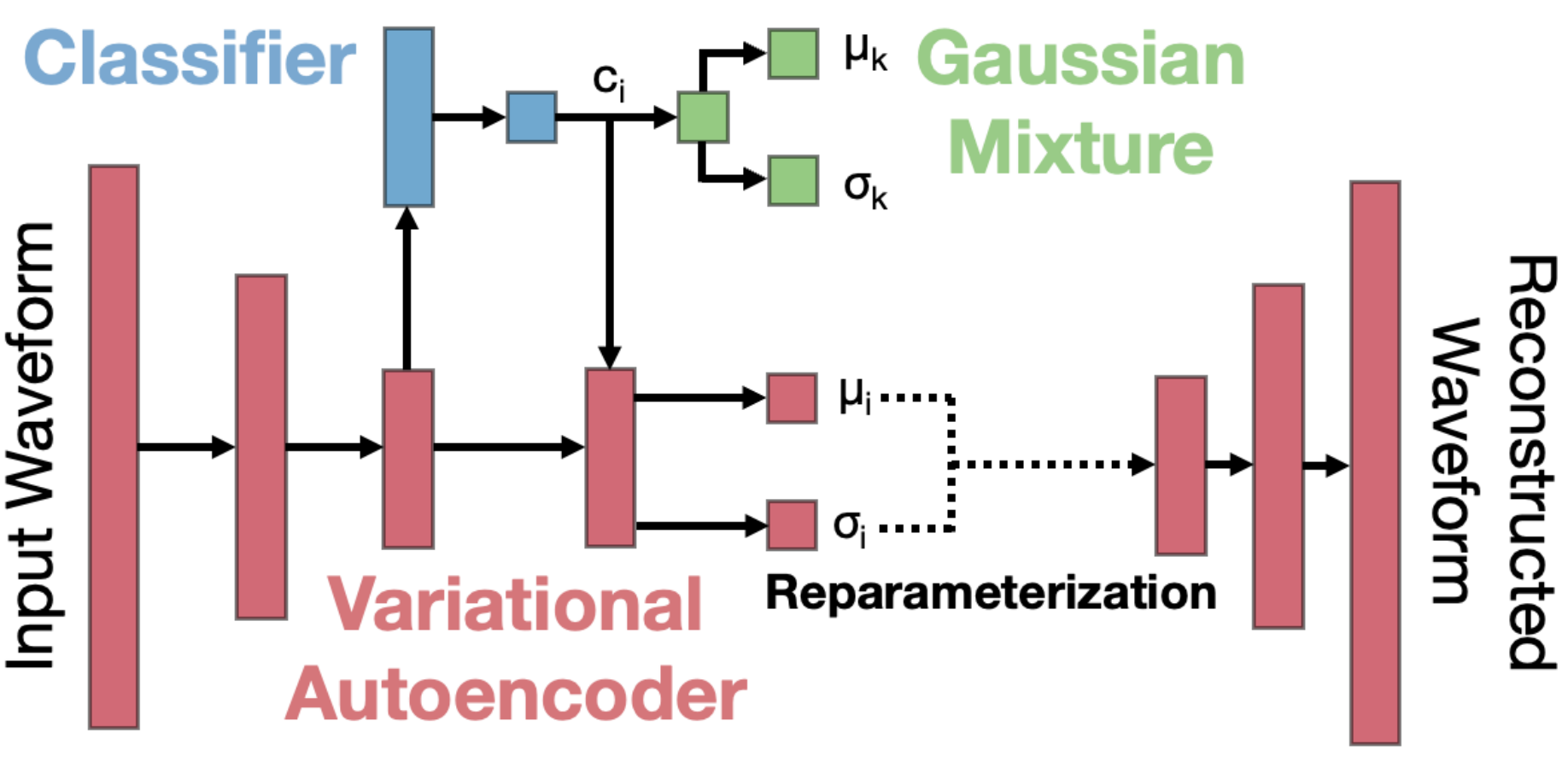}
        \caption{The architecture of the GMVAE developed for the purpose of PSD discrimination of the waveforms.}
        \label{fig:network}
    \end{minipage}\hfill
    \begin{minipage}{0.49\textwidth}
        \centering
        \includegraphics[width=\linewidth]{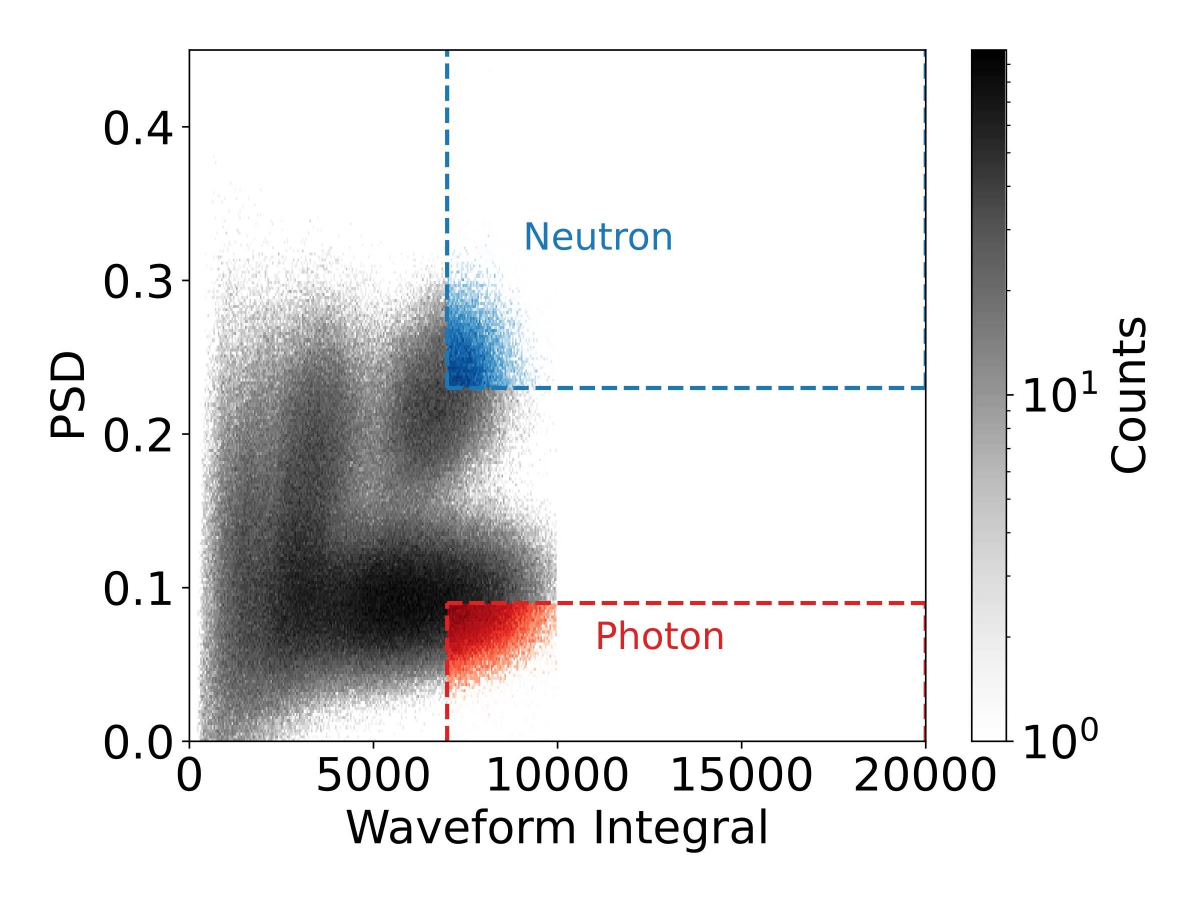}
        \caption{The data used to train the GMVAE model. The colored regions are the pre-tagged part of the data. (Refer to online plots for color).}
        \label{fig:training}
    \end{minipage}
\end{figure}

In this study, the Gaussian-Mixture Variational Auto-encoder architecture was selected with an additional classifier~\cite{Abdulaziz_2024}. The scheme of the network is shown in Figure~\ref{fig:network}, where all the different components are underlined. The auto-encoder is a generative model that is renowned for its ability to identify underlying features in an unsupervised way~\cite{Mehta_2019}, i.e.\ without the need to provide already tagged data as an input. Its main purpose is to encode the information of the incoming waveform in a latent layer and then try to decode it back. The latent space is, however, not regularized. To overcome this difficulty, the variational auto-encoder (VAE) is used instead which constrains the latent space to be Gaussian-like by applying the so-called re-parametrization trick~\cite{Mehta_2019}. The addition of a Gaussian Mixture Model~\cite{Viroli_2017} (GMM), instead, permits the VAE to sample from multiple Gaussian distributions, and effectively create a multi-modal space in the latent layer. Then, the classifier~\cite{Mehta_2019} allows for a more efficient selection of the Gaussian component and gives an instant tag for analysis purposes. In summary, the VAE permits to extract features from the waveforms and places a constraint by trying to reconstruct them; the classifier handles the tagging of each event; and the GMM selects the proper Gaussian component connected to each tag, to produce latent features where each class lies in distinct regions of the latent space.

To train the network, a subset of the dataset was prepared by taking 10,000 samples from each experimental run (140,000 samples total). Similar to~\cite{Abdulaziz_2024}, we opted for a semi-supervised method, where a small fraction of data was already tagged (using the traditional PSD approach) and the Binary Cross-Entropy loss between the network output and the tag was added to the total loss, which was defined as~\cite{Abdulaziz_2024}:

\begin{equation}
    \mathcal{L} = \mathcal{L}_{\textup{rec}} + \omega \mathcal{L}_{\textup{kld}} + \gamma \mathcal{L}_{\textup{label}}  + \theta \mathcal{L}_{\textup{triplet}} 
\end{equation}

where $\mathcal{L}_{\textup{rec}}$ is the reconstruction loss, $\mathcal{L}_{\textup{kld}}$ is the KL Divergence~\cite{Mehta_2019}, $\mathcal{L}_{\textup{label}}$ is the Binary Cross-Entropy, $\mathcal{L}_{\textup{triplet}}$ is the triplet embedding loss~\cite{Schroff_2015} and $\omega$, $\gamma$ and $\theta$ are the weights of each component. The last component penalizes large distances between features that have the same tag, thus helping the GMM to sample components that are clearly separated from each other. For the current model, $\omega = 100$, $\gamma = 100$ and $\theta = 1$ were selected. The training data are shown as the highlighted regions in Figure~\ref{fig:training}.

\begin{figure}[ht]
\centering
\includegraphics[width=\textwidth]{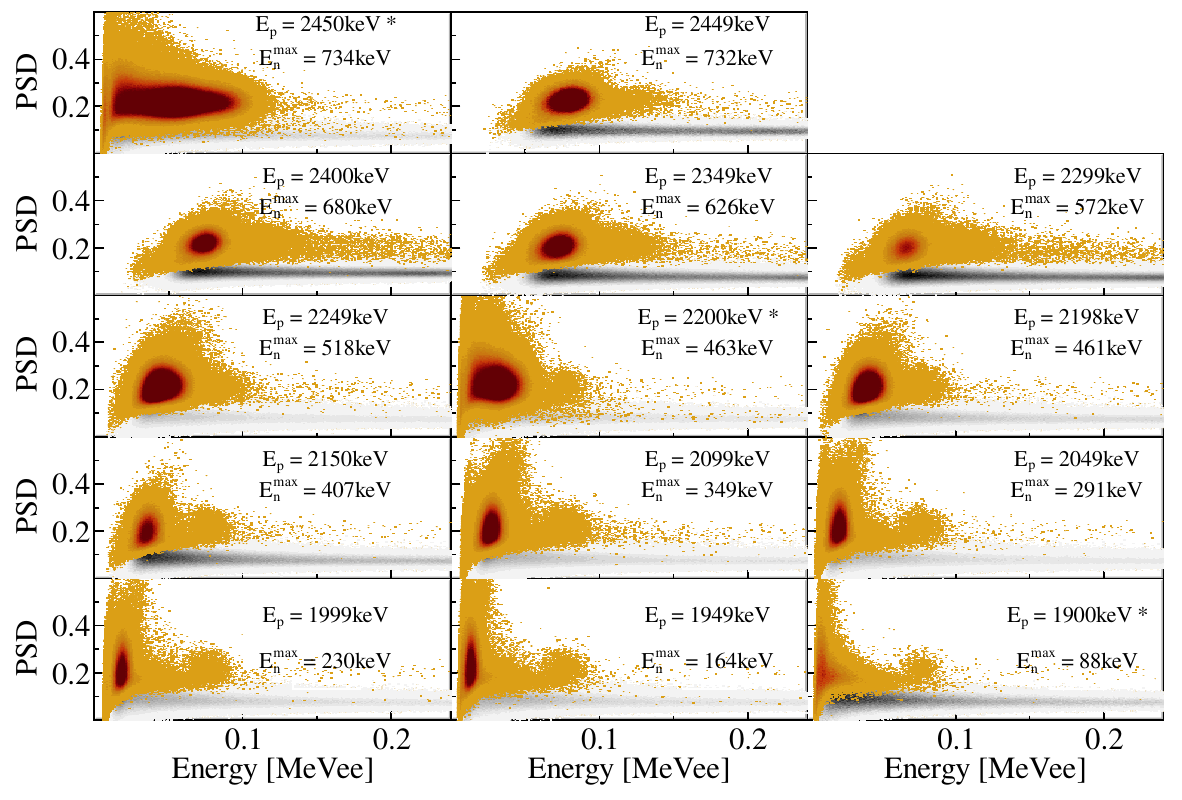}
\caption{PSD vs quenched energy using the GMVAE model results to tag the events. In black are the tagged $\gamma$-rays, and in orange are events distinguished from $\gamma$-rays. (Refer to online plots for color).}\label{fig:network:result}
\end{figure}

The training was performed with a total of 100 epochs, a batch size of 512 and a learning rate of $10^{-4}$. The dropout, i.e.\ the random deactivation of neurons at each step, was included to avoid over-fitting. Once the training was concluded, all the experimental data were passed through the model. The results are plotted in Figure~\ref{fig:network:result}, where the tags from the classifier are used to distinguish separately the $\gamma$-rays (black) from non $\gamma$-rays (orange). 

The application of the GMVAE network shows a significant advancement in low-energy event discrimination for liquid scintillator characterization. The unsupervised nature of the VAE enables the identification of subtle, non-linear features in raw waveforms, whilst the triplet embedding loss explicitly enforces class separation in the latent space. Combining these with the semi-supervised training strategy, which leverages limited tagged data to guide feature learning, permits robust discrimination even below 60\,keVee ($E_{\textrm{n}} < 475$\,keV).

\section{Prototype Array Timing Coincidence Results} \label{sec:coincidences}

\begin{figure}[ht]
\centering
\includegraphics[width=0.8\textwidth]{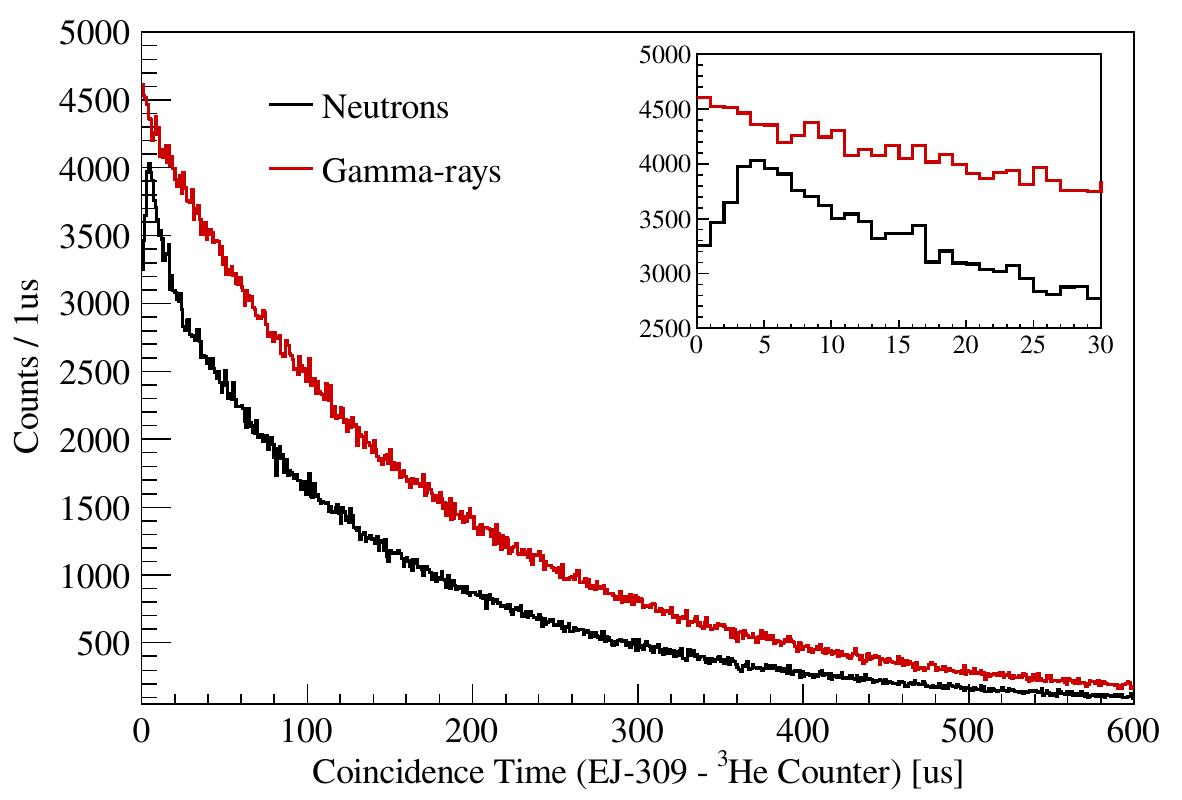}
\caption{EJ-309 - $^{3}$He counter coincidence time, selecting neutrons (black) and $\gamma$-rays (red) using the EJ-309 PSD. Inset: Zoom into region below \SI{30}{\micro\second}. (Refer to online plots for color).}\label{fig:coinTime}
\end{figure}

\begin{figure}[ht]
\centering
\includegraphics[width=0.8\textwidth]{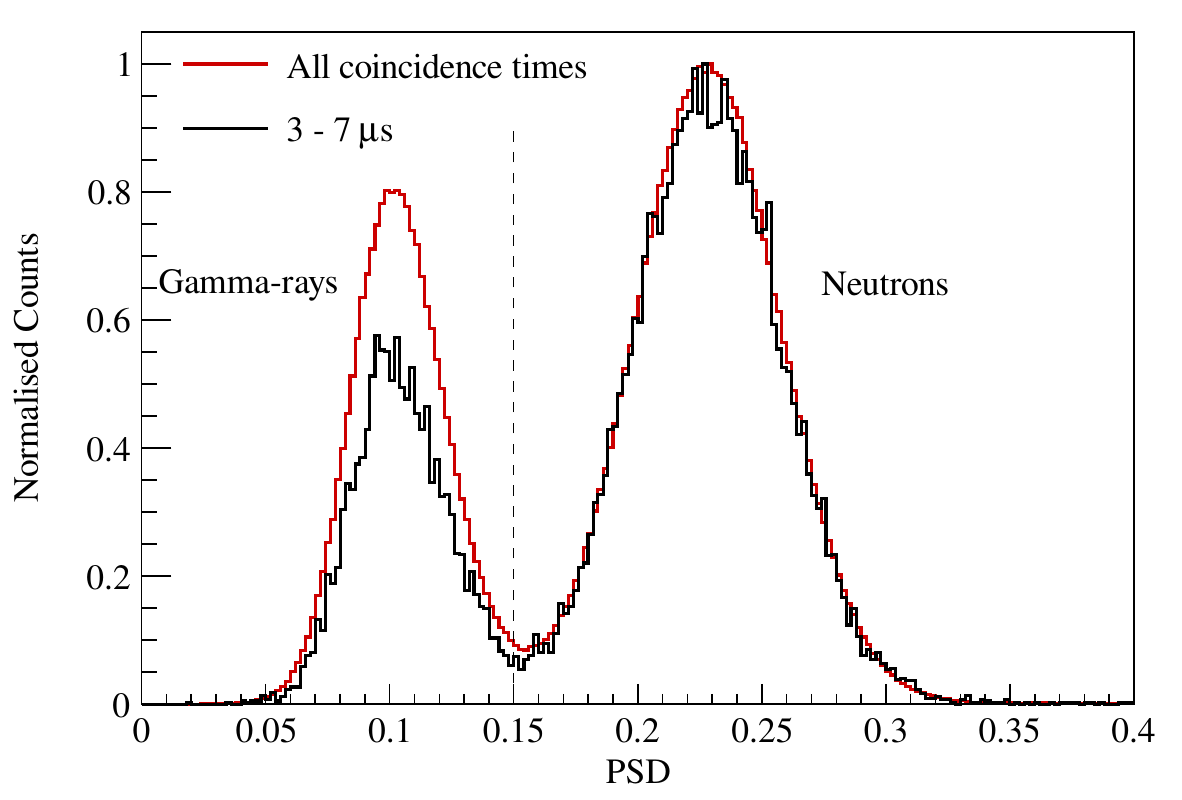}
\caption{PSD for $E_{\mathrm{p}}$ = 2449\,keV, gated on quenched energy $\leq$~150\,keVee. In red are all events, and in black are the events gated on coincidence time range 3 -- \SI{7}{\micro\second}. The peak centered around PSD = 0.23 contains the neutron signals. (Refer to online plots for color).}
\label{fig:psd_tcoinGate}
\end{figure}

The EJ-309 scintillator and $^{3}$He counter signals were all collected using the same digitizer with a common timestamp, thereby allowing the coincidence timing to be determined event-by-event. This coincidence information was used to assess the filtering capabilities of the EJ-309 - $^{3}$He counters prototype array. The coincidence time was determined by taking the difference in the timestamp of a triggered EJ-309 ($t_{\mathrm{EJ-309}}$) and its following $^{3}$He counter ($t_{\mathrm{counter}}$) event. For the $E_{\textrm{p}}$ = 2449\,keV run, this time ($t_{\mathrm{counter}}$ - $t_{\mathrm{EJ-309}}$) is plotted in Figure~\ref{fig:coinTime} where the black (red) curve shows timings for neutrons ($\gamma$-rays) as selected by gating on the traditional EJ-309 PSD. The coincidence time for neutrons peaks between 3 -- \SI{7}{\micro\second}, a feature absent for random coincidences which follow an exponential trend. The PSD spectrum gated on this coincidence figure is shown by the black histogram in Figure~\ref{fig:psd_tcoinGate} where a distinct preference is shown towards the neutron peak. This coincidence gate removes 98.9\% of the gamma-ray events and 98.4\% of the (random) neutron events, where gamma-rays are suppressed by an additional 5\% over neutrons. 

\begin{figure}[ht]
\centering
\includegraphics[width=0.8\textwidth]{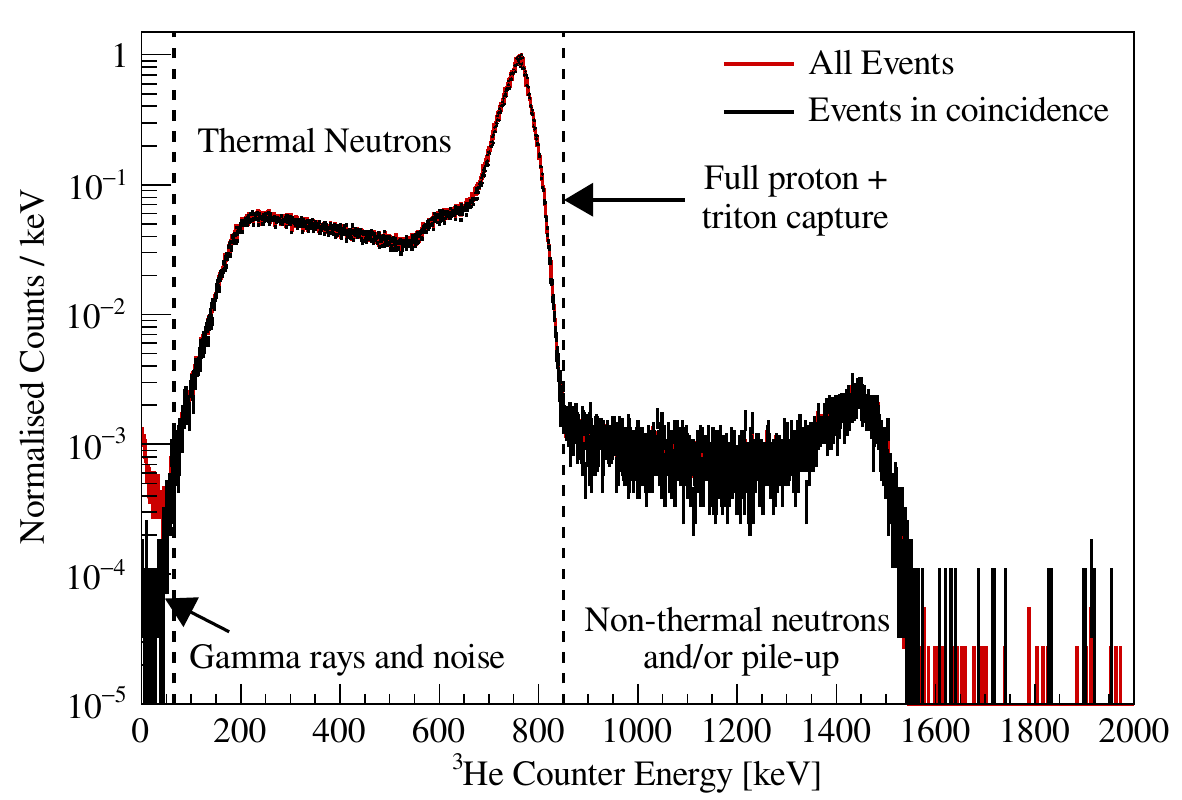}
\caption{Summed energy spectra from all six (gain-matched) $^{3}$He counters used in this experiment, normalized to peak height of the full proton + triton capture. All counter events are plotted in red, and counter events with a preceding EJ-309 coincidence event are plotted in black. (Refer to online plots for color).}\label{fig:heEnergy}
\end{figure}

The summed energy spectra from all six (gain-matched) $^{3}$He counters is provided in Figure~\ref{fig:heEnergy}. The red spectrum shows all events, whereas the black spectrum is the result of gating on any coincidence with the EJ-309. This gate reveals a factor 4.5 reduction of counts in the noise and $\gamma$-ray energy region ($E < $ 60\,keV), with an overall reduction of 48\% in all counter events. We observe clear neutron signals in the coincidence-gated spectra of Figures \ref{fig:psd_tcoinGate} and \ref{fig:heEnergy}. This highlights that the neutrons must have deposited most of their energy in the scintillator, became thermalised, and were then subsequently detected in the counters. This confirmed detection feature of the prototype array motivates expansion into the final ``SHADES" array.

\subsection{Comparison with Simulations}\label{sec:simulations}

To cross-check the observed coincidence timing, the setup was simulated using \textsc{Geant4} v11.3.0 and user-written \textsc{C++} code. The geometry included the lithium target, EJ-309 scintillator, six counters, lead blocks, and stainless steel detector supports. Simulated neutrons started in the center of the \SI{3.1}{\micro\meter} thick target, randomly distributed across a circle of diameter 10\,mm to simulate the beam spot size, and emitted in an outgoing cone of solid lab angle 8.4$^{\circ}$ towards the face of the scintillator. The neutron energy was determined via kinematical calculations~\cite{maryon_young} considering incident proton energy 2449\,keV used in the experiment and the randomly emitted neutron angle. Following \textsc{Geant4} physics lists were used: the hadron physics list \textsc{FTFP\_BERT\_HP} for the  inelastic neutron scattering, \textsc{G4ThermalNeutrons} to include a more accurate treatment of thermal neutron interactions with hydrogen (i.e. in the EJ-309), and \textsc{G4StoppingPhysics} to describe energy loss and stopping of particles. 

\begin{figure}[ht]
\centering
\includegraphics[width=0.8\textwidth]{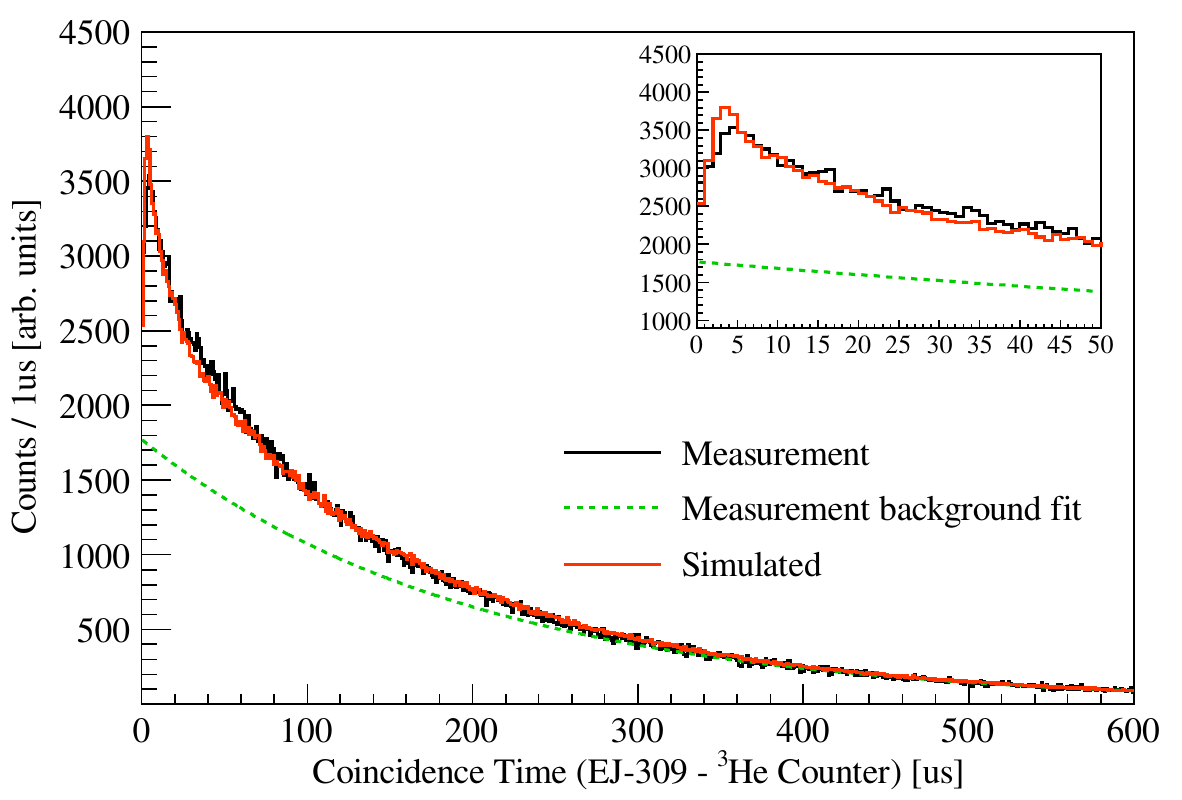}
\caption{EJ-309 - $^{3}$He counter coincidence time for neutrons from measurement in black, exponential background fitted across 400 -- \SI{600}{\micro\second} in green-dashed, and simulation in red (scaled to measured counts at \SI{5}{\micro\second} and smeared by \SI{1.2}{\micro\second}). Inset: zoom into region below \SI{50}{\micro\second}. (Refer to online plots for color).}\label{fig:coinTimeSim}
\end{figure}

The counter timing from the simulation was shifted by the charge-collection time for a cylindrical anode-cathode geometry as previously outlined in~\cite{Balibrea-Correa_2018}. The simulated coincidence timing is then defined like the experimental, $t_{\mathrm{counter}} - t_{\mathrm{EJ-309}}$. Figure~\ref{fig:coinTimeSim} compares the coincidence times between measurement (gated on neutron PSD) and simulation. The simulated coincidence curve was scaled to the measured as follows. An exponential background component was fitted to the measured data across 400 -- \SI{600}{\micro\second}. Such background dominates the measurement above \SI{100}{\micro\second} and is predicted to arise from cosmic, scattered, and radioactive decay neutrons present in the experimental hall. Ignoring the background component, the simulated curve was scaled to the experimental at \SI{5}{\micro\second} and a Gaussian filter with sigma \SI{1.2}{\micro\second} applied to take into account electronic components not included in the simulation. The background is expected to be mitigated for the future SHADES setup, which will benefit from the reduced background neutron flux at the deep-underground Gran Sasso laboratory, and by mounting the neutron detectors in 2-inch thick 5\% borated polyethylene shielding. 
 
\section{Conclusions}\label{sec:conclusions}

The neutron-detection characteristics of a prototype array constructed of a liquid scintillator surrounded by six $^{3}$He proportional counters was evaluated using a direct neutron beam produced via the $^{7}$Li(p,n$_{0}$)$^{7}$Be reaction at the ``FRANZ" facility. Three essential properties were studied for neutron detection experiments; the scintillator's neutron/$\gamma$-ray PSD, the scintillator-counter coincidence features, and the scintillator's moderation of thermal neutrons for capture in the counters. The EJ-309 PSD was evaluated using both a traditional charge-integration approach and a trained neural network. Whilst for high neutron energies the traditional PSD is sufficient for neutron/$\gamma$-ray discrimination, the neural network provides neutron/$\gamma$-ray tagging capabilities down to neutron energies as low as 163\,keV. The EJ-309 - $^{3}$He counter coincidence provides a suppression factor of 48\% from background neutrons entering the counters. The coincidence also reveals a distinct peak-like feature for neutrons around 3 -- \SI{7}{\micro\second}, found to be reproducible by Monte-Carlo simulations considering the empirical charge processing times. Additionally, the counter energy spectrum gated on EJ-309 events emphasizes the EJ-309 capability to thermalize neutrons for prompt capture by the $^{3}$He counters. These compounding features of the prototype array led to the construction and development of the full-scale version titled ``SHADES", currently undergoing measurements at the LNGS IBF facility. Here, the SHADES array is working to directly measure the $^{22}$Ne($\alpha$,n)$^{25}$Mg reaction cross-section at energies of astrophysical interest in RGB and AGB stars for the main and weak $s$-processes.


\section{Supplementary information}

The raw data is provided in the INFN open access repository: \\
Repository 1 of 2: https://doi.org/10.15161/oar.it/kmkvr-qdj73 \\
Repository 2 of 2: https://doi.org/10.15161/oar.it/bhkhq-8ar22

A copy of the Geant4 simulation code is provided on the public git repository: https://baltig.infn.it/LUNA/frankfurt-sim/

\section{Acknowledgments}

This research was supported by the European Union through ERC-StG 2019 \#852016 ``SHADES" and the Horizon 2020 research and innovation programme grant agreement No 101008324 (ChETEC-INFRA). We thank the beam production team at the ``FRANZ" facility for providing the neutron beam.

\Large References

\bibliographystyle{unsrt}
\bibliography{bibliography}

\end{document}